\documentclass{jors}

\definecolor{mygray}{gray}{0.6}

\usepackage{amsfonts,amsmath,amssymb}		
\usepackage{bm}
\usepackage{color}
\usepackage{float}
\usepackage{graphicx}
\usepackage[numbers,sort&compress]{natbib}
\usepackage{upgreek}

\usepackage{url}
\RequirePackage{doi}
\usepackage{hyperref}

\newcommand{\ie}{{i.e.}}
\newcommand{\eg}{{e.g.}}

\newcommand{\figref}[1]{Fig.~\ref{fig:#1}}

\newcommand{\figlab}[1]{\label{fig:#1}}

\newcommand{\seclab}[1]{\label{sec:#1}}

\newcommand{\Cm}{C_\mathrm{m}}

\begin{document}

{\bf Software paper for submission to the Journal of Open Research Software} \\

To complete this template, please replace the blue text with your own. The paper has three main sections: (1) Overview; (2) Availability; (3) Reuse potential. \\

Please submit the completed paper to: editor.jors@ubiquitypress.com

\rule{\textwidth}{1pt}

\section*{(1) Overview}

\vspace{0.5cm}

\section*{Title}
\emph{DefocusTracker}: A modular toolbox for defocusing-based, single-camera, 3D particle tracking

\section*{Paper Authors}

1. Barnkob, Rune;\\
2. Rossi, Massimiliano;

\section*{Paper Author Roles and Affiliations}
1. Heinz-Nixdorf-Chair of Biomedical Electronics, Department of Electrical and Computer Engineering, Technical University of Munich, Center for Translational Cancer Research (TranslaTUM), Munich, Germany\\
2. Department of Physics, Technical University of Denmark, DTU Physics Building 309, DK-2800 Kongens Lyngby, Denmark

\section*{Abstract}

The need for single-camera 3D particle tracking methods is growing, among others, due to the increasing focus in biomedical research often relying on single-plane microscopy imaging. Defocusing-based methods are ideal for a wide-spread use as they rely on basic microscopy imaging rather than requiring additional non-standard optics. However, a wide-spread use has been limited by the lack of accessible and easy-to-use software. \emph{DefocusTracker} is an open-source toolbox based on the universal principles of General Defocusing Particle Tracking (GDPT) relying solely on a reference look-up table and image recognition to connect a particle's image and its respective out-of-plane depth coordinate. The toolbox is built in a modular fashion, allowing for easy addition of new image recognition methods, while maintaining the same workflow and external user interface. \emph{DefocusTracker} is implemented in MATLAB, while a parallel implementation in Python is in the preparation.

\section*{Keywords}

particle tracking; velocimetry;
PTV;
general defocusing particle tracking;
fluid dynamics;
MATLAB;
Python

\section*{Introduction}
\seclab{intro}

The use of single-camera 3D particle tracking analysis is receiving increasing interest, among others, due to the rapid development of bio-engineering and biomedical sciences where single-access imaging, such as with microscopes, is a standard research tool~\cite{taute2015high,van2020gradient}. For this, methods based on the principle of particle image defocusing are particularly attractive as no special optics or cameras are required, and have potential for wide-spread use. 
However, until now most of the software for defocused-based particle tracking has been developed in-house for private use of research groups, and there are only few examples of user-friendly software that can be accessible to a larger audience, including researchers outside the engineering or computer-science community. One example is GDPTlab, a MATLAB GUI implementation written by the authors and released in 2015 \cite{GDPTlab}. GDPTlab was used by few research groups and cited in several peer-reviewed journals (see also Reuse potential section). GDPTlab, however, was not distributed under an open-source license, thus its potential for collaboration and expansion was limited by that.  

\indent
To accommodate this need, we developed \emph{DefocusTracker}, which is a modular and open-source toolbox for defocusing-based 3D particle tracking. \emph{DefocusTracker} uses different architecture and functions and is not compatible with GDPTlab, however they are based on the same method, namely the General Defocusing Particle Tracking (GDPT). GDPT relies on a reference look-up table, with known defocused particle images and depth positions, and on a image recognition method, that matches target and reference particle images~\citep{barnkob2015general,taute2015high,barnkob2020general}. Following machine learning terminology, we will refer here to the look-up table as the training set, whereas the image recognition method will be part of the chosen calibration model. The GDPT principle is shown in \figref{architecture-general}. Panel (a) shows the creation and training of a calibration model through calibration/training images of defocused particle images with known 3D particle positions. In Panel (b) the trained model is used to reconstruct the 3D particle positions from the particles' defocused images in 2D measurement images. For more details on the GDPT method, experiments, and uncertainty assessment, we refer to Refs.~\citenum{barnkob2015general,barnkob2020general,GDPTlab}.

\indent \emph{DefocusTracker} is built in a modular fashion to allow for a continuous addition of state-of-the-art image recognition methods, \eg\ methods based on convolutional neural networks and deep learning, while maintaining general method-independent features. The current implemented method, referred to as Method 1, is based on the normalized cross-correlation function and described in Ref. ~\citenum{rossi2020toward}. \emph{DefocusTracker} is implemented in MATLAB, while a parallel implementation in Python is in the preparation. \emph{DefocusTracker} is accompanied by a website,  \url{https://defocustracking.com/}, to facilitate the research community with a platform for sharing of user guides, experiences, and applications, as well as for  data for training and validation. 

\begin{figure*}[t!]
    \includegraphics[width=1.0\textwidth]{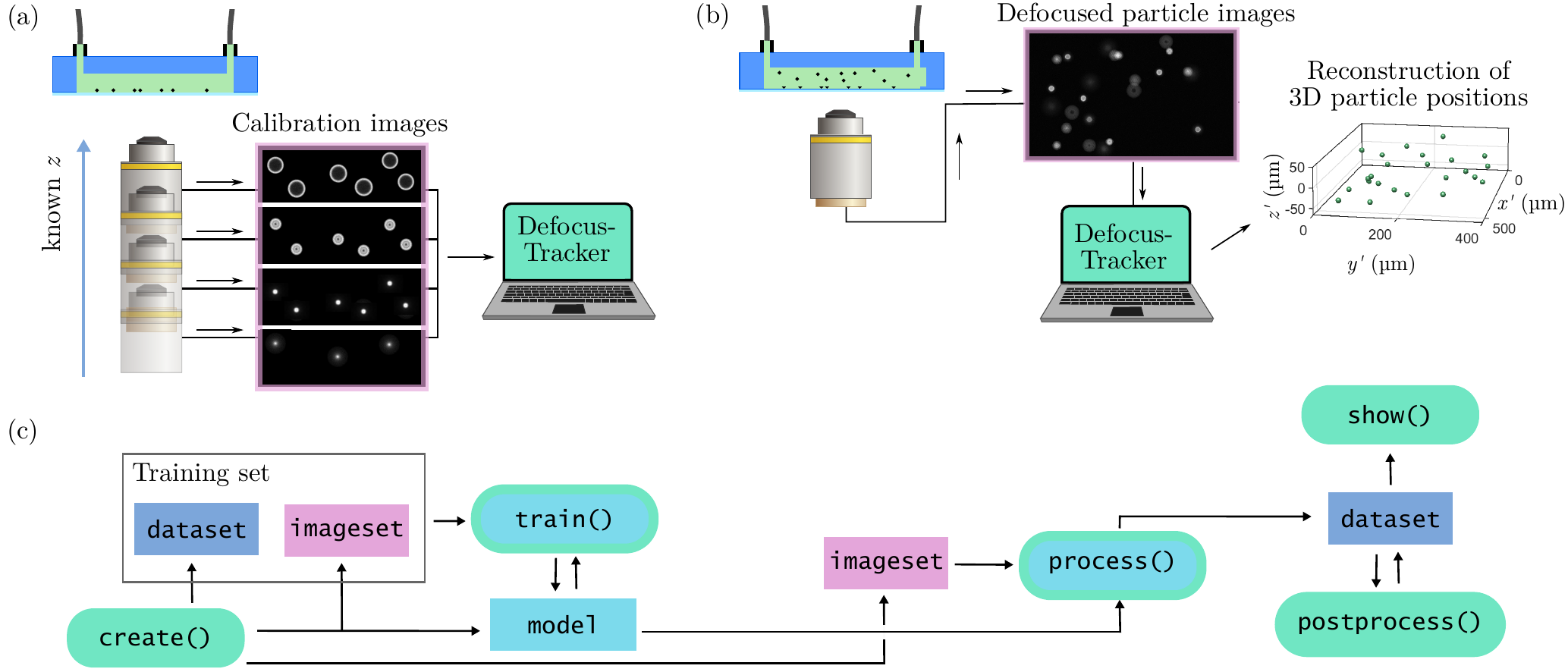}
    \centering
	\caption{\emph{DefocusTracker} working principle and general architecture. 	(a) A set of calibration images with known 3D particle positions are used to (b) determine the unknown 3D positions of particles through the comparison of their defocused particle images. (c) General architecture and workflow of the \emph{DefocusTracker} toolbox. The toolbox uses three types of data structures (rectangles) and five main functions (round shapes, green). The toolbox is modular, allowing for addition and use of different models for the image processing and particle tracking.}
	\figlab{architecture-general}
\end{figure*}

\vspace{5mm}
\section*{Implementation and architecture}

\vspace{5mm}
\textbf{General architecture}

The general architecture and workflow is shown in \figref{architecture-general}(c). The toolbox is based on three types of data structures and five primary functions for their creation, processing, and manipulation:\\

\underline{Data structures:}
\begin{description}
\item[\texttt{imageset}] Link and description to a set of images, \ie\ image paths, number of images, and image type.
\item[\texttt{dataset}] Particle data, \ie\ 3D spatial positions and displacements, trajectories, connection to image frames, and detection accuracy estimation.
\item[\texttt{model}] Data and settings required to process the images, \ie\ method-specific parameters as well as training and processing settings.
\end{description}

\underline{Functions:}
\begin{description}
\item[\texttt{create()}] Creates a data structure.
\item[\texttt{show()}] Opens GUIs to inspect data structures.
\item[\texttt{train()}] Trains a \texttt{model} on a specific training set (\texttt{imageset}+\texttt{dataset}).
\item[\texttt{process()}] Processes an \texttt{imageset} using a given \texttt{model}.
\item[\texttt{postprocess()}] Manipulates a \texttt{dataset}, \eg\ merge datasets, apply scaling, remove outliers, perform particle tracking, filter trajectories, and estimate uncertainties.
\end{description}

\underline{Workflow:}\\
As seen in \figref{architecture-general}(c), a typical workflow starts with the creation of a \texttt{model} using the \texttt{create()} function. Each \texttt{model} refers to a specific method (e.g. Method 1) and it is at first created with some default values. The \texttt{model} is trained using the \texttt{train()} function by feeding a training set as input. A training set consists of a \texttt{dataset} and an \texttt{imageset} (made with \texttt{create()}), corresponding to a set of images containing one or more particles of known 3D position. As illustrated in \figref{architecture-general}(a), such a training set can be obtained experimentally by taking subsequent images of particles displaced at known positions, e.g. by observing particles sedimented on a microchannel bottom, while taking images at known objective distances using a focusing stage. If only a subset of the calibration particles are used for a training, the remaining particles can be used as validation to make a pre-measurement uncertainty estimation.

\indent With a trained \texttt{model} as input, the \texttt{process()} function can take one or more measurement images in an \texttt{imageset} and output a \texttt{dataset} containing the measured 3D particle positions. The \texttt{dataset} can be further manipulated via the \texttt{postprocess()} function for purposes such as outlier removal or trajectory smoothing. Throughout the entire workflow, the \texttt{show()} function can be used to visualize and inspect the data structures.\\

\underline{Modularity and Method 1:}\\
The \emph{DefocusTracker} toolbox is modular in the sense that a \texttt{model} can be created based on different methods for the image recognition. If a new method is added to the toolbox, the data structures, primary functions, and workflow will remain the same. The current implementation provides only one method for the creation of a \texttt{model}, namely Method 1. For a full description of Method 1, we refer to \cite{rossi2020toward}. Briefly, Method 1 is based on training images of a single particle (the so-called calibration stack) and uses the normalized cross-correlation function for image recognition. The normalized cross-correlation is used to rate the similarity between a target particle image and the calibration stack images, using its maximum peak value as the similarity coefficient, referred to as $\Cm$. The values of $\Cm$ can range from 0 to 1, with 1 corresponding to a perfect match between the target image and a calibration image.

\begin{figure*}[t!]
    \includegraphics[width=1.0\textwidth]{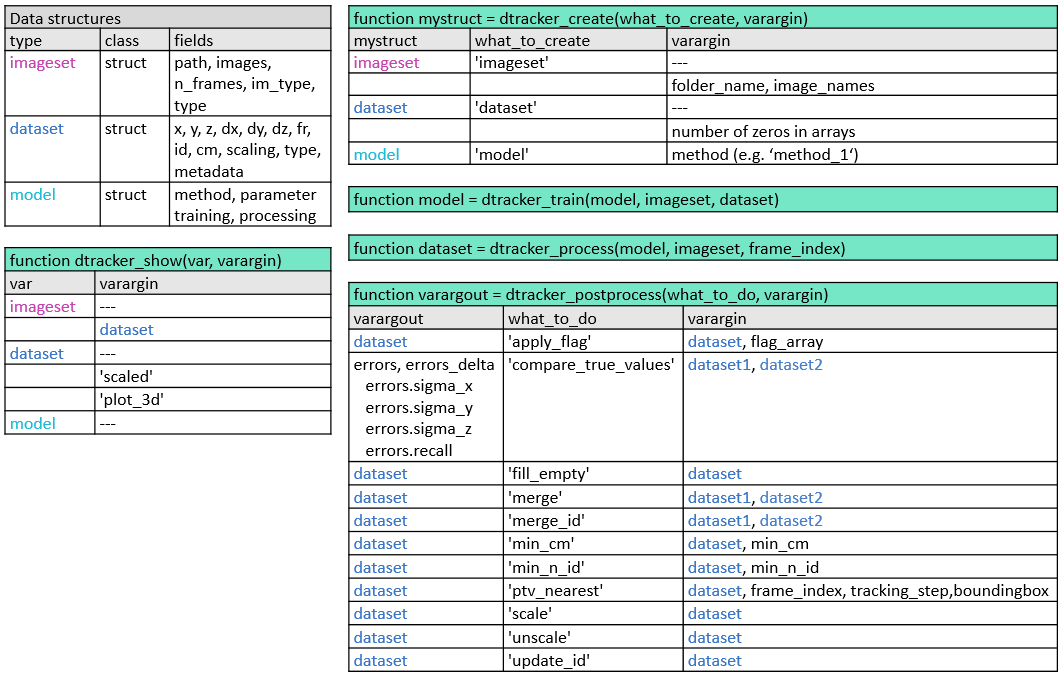}
    \centering
	\caption{Overview of data structures and functions in the \emph{DefocusTracker} MATLAB implementation, following the general toolbox architecture shown in \figref{architecture-general}.}
	\figlab{architecture-matlab}
\end{figure*}

\vspace{5mm}
\textbf{MATLAB implementation}

\emph{DefocusTracker} is implemented in MATLAB and additionally requires the image processing, curve fitting, and statistics toolboxes. The implementation is script-based with certain features using GUI-based pop-up windows for visualization. The data structures are so-called MATLAB structs, namely structure arrays where data is grouped using containers called fields. The data in a field is accessed using dot notation of the form \texttt{structName.fieldName}, \eg\ the path of an \texttt{imageset} is called with \texttt{imageset.path}. The primary toolbox functions follow the form of standard MATLAB functions and are named as \texttt{dtracker\_functionName()}, \eg\ \texttt{dtracker\_create()}. A full overview of the data structures and functions are shown in \figref{architecture-matlab}. An exemplary practical application of \emph{DefocusTracker}, including a script and few results, is presented in \figref{WTE2}. More details about this application are given in the Quality control section.\\

\vspace{5mm}
\textbf{Python implementation}

A Python implementation of \emph{DefocusTracker} is planned and under development and it will the lines of the MATLAB implementation. The data structures will be implemented using Python dictionaries, whereas the functions will be part of the module \texttt{dtracker}. Following the above example, the path of an \texttt{imageset} will be called in Python with \texttt{imageset['path']}, whereas the \texttt{create()} function will be called as \texttt{dtracker.create()}. 

\indent
The Python implementation is available on the Gitlab repository: \url{https://gitlab.com/defocustracking/defocustracker-python}. Updates and release information can be followed on \url{https://defocustracking.com}.  

\begin{figure*}[t!]
    \includegraphics[width=1.0\textwidth]{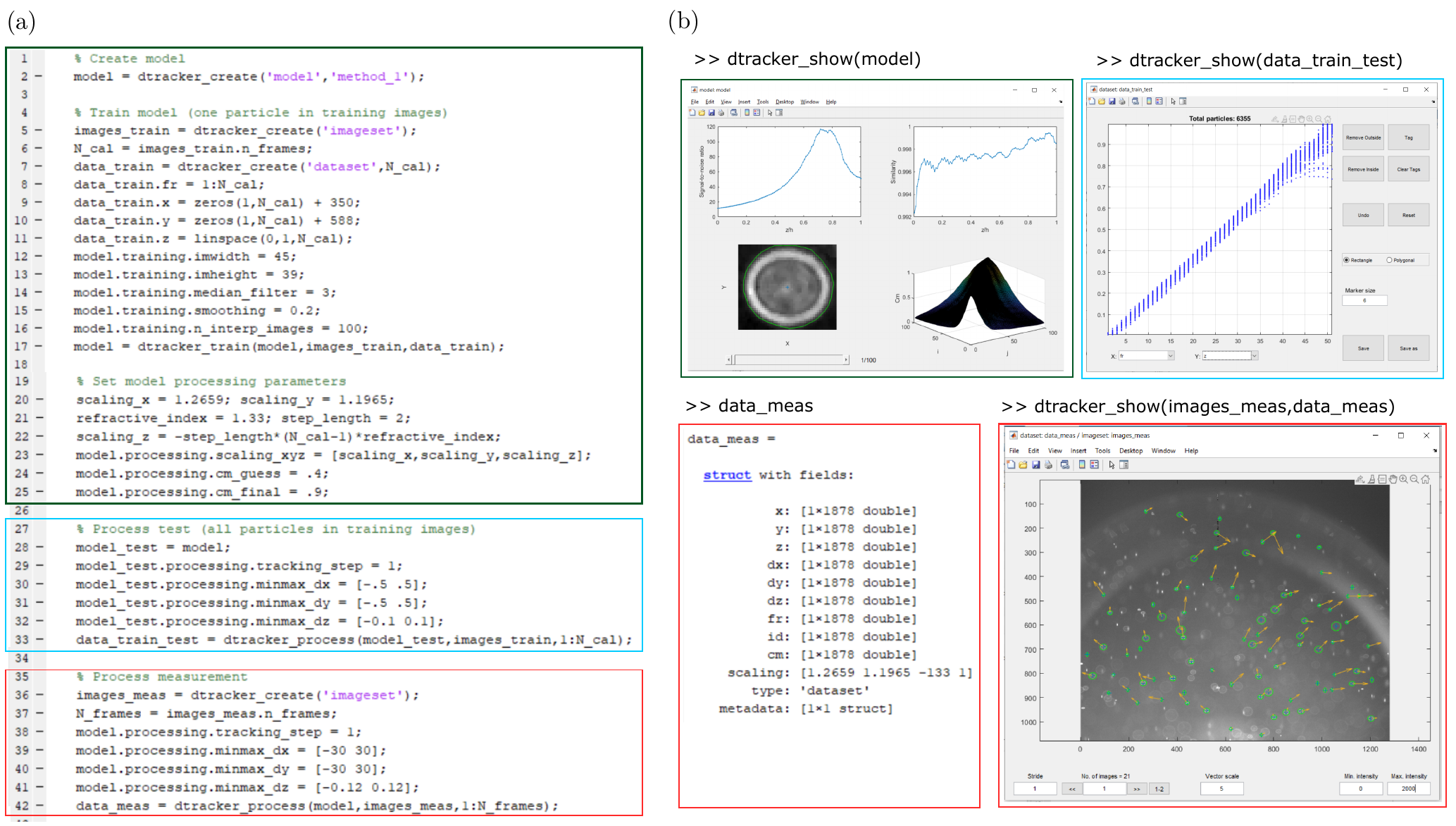}
    \centering
	\caption{Example workflow of the toolbox MATLAB implementation. The example workflow is based on part of the provided Work-Through Example 2 that takes the user through the processing and analysis of particle trajectories inside an evaporating droplet~\cite{rossi2019interfacial}. (a) Illustration of the code used to create and train a model (green frame), to process image test data (blue frame), and to process measurement images (red frame). (b) Illustration of the implemented pop-up GUIs used to visualize and inspect the data structures.}
	\figlab{WTE2}
\end{figure*}

\section*{Quality control}

The MATLAB toolbox has been tested functionally on Windows 10 with MATLAB releases R2018b and 2020a. The toolbox performance has been tested and investigated extensively in two recent publications~\cite{barnkob2020general,rossi2020toward}. In \citet{barnkob2020general}, guidelines for assessing the uncertainty of GDPT analyses were given. Synthetic images were used to test the toolbox in terms of measurement uncertainty and relative number of measured particles as a function of image signal-to-noise ratio, particle image concentration, and variations in image intensity. In \citet{rossi2020toward}, different toolbox settings were tested on synthetic and experimental images to outline the measurement uncertainties, detection rates, and processing times. The results were benchmarked against the GDPTlab software~\citep{GDPTlab}, which has been extensively-tested and used in high-impact research publications, see more in the Section Reuse potential.

\indent The MATLAB toolbox contains two work-through examples (WTE1 and WTE2) that serve as tutorials to get new users quickly started, but also as test scripts in case new functionalities or methods are added. The scripts of each WTE are included in the \emph{DefocusTracker} package, while the relative datasets can be acquired via \url{https://defocustracking.com/datasets/}.  WTE1 is based on synthetic images and gives a first introduction to the basic building blocks of the toolbox. The use of synthetic images allow for an exact estimation of the uncertainty using the postprocessing method \texttt{'compare\_true\_values'}.
WTE2 is based on a state-of-the art microfluidic experiment, namely the 3D flow inside an evaporating droplet~\citep{rossi2019interfacial}, and guides the user toward a more advanced use of \emph{DefocusTracker}, including postprocessing and bias correction. We report in \figref{WTE2} a shortened version of the WTE2 script, including few screenshots of GUI panels obtained with the \texttt{dtracker\_show()} function. For the full commented version we refer to the script \texttt{Work\_through\_ex2.m}. As the community grows, we expect that more WTEs will be added by the users and developers.

\section*{(2) Availability}
\vspace{0.5cm}
\section*{Operating system}

Windows, UNIX/Linux, Macintosh (and any operating system supporting MATLAB).

\section*{Programming language}

MATLAB 9.4.0 (R2018a), upward compatible. 

\section*{Additional system requirements}

N/A

\section*{Dependencies}

The MATLAB implementation requires the additional MATLAB toolboxes:\\ `curve\_fitting\_toolbox`, `image\_toolbox`, `statistics\_toolbox`

\section*{List of contributors}

N/A

\section*{Software location:}

{\bf Archive} 

\begin{description}[noitemsep,topsep=0pt]
	\item[Name:] Gitlab (master)
    \item[Persistent identifier:] %
	https://gitlab.com/defocustracking/defocustracker-matlab/-/releases/v1.0.2
	\item[Licence:] MIT
	\item[Publisher:]  Massimiliano Rossi and Rune Barnkob
	\item[Version published:] 1.0.2
	\item[Date published:] 01/10/2020
\end{description}

{\bf Code repository}

\begin{description}[noitemsep,topsep=0pt]
	\item[Name:] Gitlab (develop)
	\item[Persistent identifier:] https://gitlab.com/defocustracking/defocustracker-matlab
	\item[Licence:] MIT
	\item[Date published:] 01/10/2020
\end{description}

\section*{Language}

English

\section*{(3) Reuse potential}

The analysis of particle positions, velocities, and trajectories is an integral part of many research disciplines. This includes analyses in 2D as well as in 3D, and with the rapid growth in fields relying on microscopy, such single-camera methods can provide unique and important information. One example is within the field of microfluidics where high control of fluid flow and externally-applied forces is becoming an important tool in biomedical research and applications. Here, 3D detection and tracking of particles and cells can provide the necessary information needed for optimization, standardization, and real-time inspection and control~\cite{sitters2015acoustic,sapudom2017quantitative}.

\indent GDPT has shown to be an excellent candidate for a wide-spread technique as is a simple and universal defocusing-based method and requires no special optics and can be used in standard microscope setups. Here, the development of free, accessible, user-friendly, and accurate tools can greatly enhance the practicability and availability of the method. One example is the MATLAB implementation GDPTlab (also by the authors), which has been distributed to researchers since 2015 and has proven its value in a number of research projects including work in journals such as Proceedings of the National Academy of Sciences, Physical Review Letters, and Scientific Reports~\citep{rossi2017kinematics,barnkob2018acoustically,volk2018size,Qiu2019,karlsen2018acoustic,liu2019investigation,qiu2020particle,van2020gradient,bode2020microparticle,westerbeek2020reduction}. Note that most of this research were done in laboratories with no previous experience of 3D single-camera particle tracking prior to the use of GDPTlab. GDPTlab has thus shown the huge potential such methods hold, if free and user-friendly implementations are available. Though GDPTlab has enabled many researchers to get started using GDPT, it has unfortunately not been fully accessible as an open-source project, limiting its further development and adaptation by the community. \emph{DefocusTracker} fills this need as it is fully open-source. The toolbox contains a readme-file and work-through examples for users to get quickly started.

\vspace{5mm}
\textbf{Modification and support}

\emph{DefocusTracker} is set up in a versatile and modular fashion allowing for easy expansions and improvements, such as extensions of custom functionalities and features, \eg\ using MATLAB's GUI editor and pre-built functions. In the Python implementation, such expansions could involve the use of popular libraries for data analysis and machine learning, such as SciKit, Keras, and TensorFlow. \emph{DefocusTracker} is supported by \url{https://defocustracking.com/} which is an online platform created to assist the development and support of \emph{DefocusTracker} as well as to facilitate the research community with a place for sharing of data and experiences related to single-camera 3D particle tracking. The platform contains several forums, \eg\ for new developers to request access and for users to ask the community for support.



\section*{Funding statement}

The research leading to these results has received funding from the European Union’s Horizon 2020 research and innovation programme under the Marie Sklodowska-Curie grant agreement no. 713683 (COFUNDfellows-DTU).

\section*{Competing interests}

The authors declare that they have no competing interests.

\vspace{2cm}

\rule{\textwidth}{1pt}

{ \bf Copyright Notice} \\
Authors who publish with this journal agree to the following terms: \\

Authors retain copyright and grant the journal right of first publication with the work simultaneously licensed under a  \href{http://creativecommons.org/licenses/by/3.0/}{Creative Commons Attribution License} that allows others to share the work with an acknowledgement of the work's authorship and initial publication in this journal. \\

Authors are able to enter into separate, additional contractual arrangements for the non-exclusive distribution of the journal's published version of the work (e.g., post it to an institutional repository or publish it in a book), with an acknowledgement of its initial publication in this journal. \\

By submitting this paper you agree to the terms of this Copyright Notice, which will apply to this submission if and when it is published by this journal.

\end{document}